\documentclass{elsart3-1}

 \usepackage{graphicx}

\usepackage{amssymb}

\usepackage[english,francais]{babel}

\newtheorem{e-proposition}[theorem]{Proposition}

\newtheorem{e-definition}[theorem]{Definition\rm}

\renewcommand{\theequation}{\arabic{equation}}
\def\og{\leavevmode\raise.3ex\hbox{$\scriptscriptstyle\langle\!\langle$~}}
\def\fg{\leavevmode\raise.3ex\hbox{~$\!\scriptscriptstyle\,\rangle\!\rangle$}}

\usepackage[T1]{fontenc}
\usepackage{xfrac}
\graphicspath{ {./figs/} }
\usepackage{amsmath}
\usepackage{FC-def}

\begin{document}

\begin{frontmatter}


\title{An objective perspective for classic flow classification criteria}


\selectlanguage{english}
\author[ad1]{R. S. Martins},
\ead{ramon.martins@polytech-lille.fr}
\author[ad1]{A. S. Pereira},
\ead{anselmo.pereira@polytech-lille.fr}
\author[ad1]{G. Mompean},
\ead{gilmar.mompean@polytech-lille.fr}
\author[ad1]{L. Thais},
\ead{laurent.thais@polytech-lille.fr}
\author[ad2]{R. L. Thompson}
\ead{rthompson@id.uff.br}

\address[ad1]{Université de Lille 1 - Sciences et Technologies, Polytech'Lille, and Laboratoire de Mécanique de Lille (LML), UMR-CNRS 8107, Cité Scientifique, 59655 Villeneuve d'Ascq, France.}
\address[ad2]{Laboratório de Mecânica Teórica Aplicada (LMTA), Department of Mechanical Engineering, Universidade Federal Fluminense, Rua Passo da Pátria 156, Niterói, RJ 24210-240, Brazil.}


\medskip
\begin{center}
{\small Received *****; accepted after revision +++++\\
Presented by £££££}
\end{center}

\begin{abstract}
Four classic criteria used to the classification of complex flows are discussed here. These criteria are useful to identify regions of the flow related to shear, elongation or rigid-body motion. These usual criteria, namely $\q$, $\d$, $\ltwo$ and $\lcrlci$, use the fluid's rate-of-rotation tensor, which is known to vary with respect to a reference frame. The advantages of using objective (invariant with respect to a general transformation on the reference frame) criteria are discussed in the present work. In this connection, we construct versions of classic criteria replacing standard vorticity, a non-objective quantity, by effective vorticity, a rate of rotation tensor with respect to the angular velocity of the eigenvectors of the strain rate tensor. The classic criteria and their corresponding objective versions are applied to classify two complex flows: the transient ABC flow and the flow through the abrupt 4:1 contraction. It is shown that the objective versions of the criteria provide richer information on the kinematics of the flow.
{\it To cite this article: R. S. Martins, A. S. Pereira, G. Mompean, L. Thais, R. L. Thompson, C. R.
Mécanique xxx (2015).}

\vskip 0.5\baselineskip

\selectlanguage{francais}
\noindent{\bf R\'esum\'e}
\vskip 0.5\baselineskip
\noindent
{\bf Une perspective objective pour des critères classiques de classification d'écoulements.} 
Quatre critères classiques utilisés pour la classification des écoulements complexes sont étudiés ici. Ces critères sont utiles pour identifier les régions de l'écoulement liées au cisaillement, à l'extention ou au mouvement de corps rigide. Ces critères habituels, à savoir $\q$, $\d$, $\ltwo$ and $\lcrlci$, utilisent le tenseur taux de rotation du fluide qui est connue pour varier par rapport au système de référence. Les avantages d'utiliser des critères objectives (invariants par rapport à une transformation générale pour un système de référence) sont discutées dans le present travail. À cet égard, nous construisons des versions des critères classiques en remplaçant la vorticité standard, une quantité non-objective, par la vorticité effective, un taux de rotation par rapport à la vitesse angulaire des vecteurs propres du tenseur taux de déformation. Les critères classiques et leurs versions objectives correspondantes sont appliqués pour classifier deux écoulements complexes~: l'écoulement ABC transitoire et l'ècoulement à travers une contraction brusque 4:1. Les version objectives des ces critères fournicent des informations plus riches pour la cinématique de l'écoulement.   
{\it Pour citer cet article~: R. S. Martins, A. S. Pereira, G. Mompean, L. Thais, R. L. Thompson, C. R.
Mécanique xxx (2015).}

\vskip 0.5\baselineskip
\noindent{\small{\it Mots-cl\'es~:} Objectivité; Classification~d'écoulements; Écoulement~ABC; Contraction; Identification~de~tourbillons}}
\end{abstract}
\end{frontmatter}


\selectlanguage{english}
\section{Introduction}\label{sec:intro}
In Fluid Mechanics, flow visualization is an important subject, since fundamental aspects of the flow can be captured by observation. Post-processing Computational Fluid Dynamics (CFD) data is also a field that makes important contributions for the understanding of the flow. Complex flows exhibit different kinds of motion that depend on position and {\bf time}. In these flows, it is common to find swirling motions in different parts of the domain. In order to locate and visualize these regions, a criterion of vortex identification is generally used to see the manifestation of the rotation character of the flow.

However, the concept of a \emph{vortex} is still cause for dissension within the scientific community. As a consequence, there a several criteria available in the literature that are used to identify rotation structures in the flow. In other words, there is no quantity, in the mathematical sense, that is consensually accepted in the literature as a definition for a vortex. Some of the non-consensual issues that are present in this context are if the vortex is an Eulerian or a Lagrangian entity and if it should be defined in a kinematic or in a dynamical basis. 

Comparisons among the different criteria are still a subject of investigation (e.g. \cite{Pierce-13}). An important point raised by Haller \cite{Haller-05} is the requirement that a vortex should be an Eulerian invariant entity, i.e. invariant under arbitrary changes of the reference frame. This requirement affects the vortex concept, since, before that work, only the Galilean invariance was invoked to define a vortex \cite{Jeong-95}. We can stress here that the classic vortex definitions, such as the $\q$-criterion by Hunt \etal \cite{Hunt-88}, the $\d$-criterion by Chong \etal \cite{Chong-90}, the $\ltwo$-criterion by Jeong and Hussain \cite{Jeong-95} and the $\lcrlci$-criterion by Chakraborty \etal \cite{Chakraborty-05}, enjoy only Galilean invariance (i.e. they are invariant to constant velocity translating frames).



The arguments to adopt an objective criterion for vortex identification are the following. First, if one observer identifies a certain region as being a vortex while, for another one, this region is not a vortex, there is no reason to privilege the verdict stated by one observer with respect to the other. Secondly, we have to have in mind the advantages of building a criterion for vortex identification. One clear purpose of identifying a region as being a vortex is to connect the rotational character of the flow with another phenomenon besides the flow itself. It is consensual that processes like: the convection in a heat transfer problem, the degree of mixture of different fluids, the percentage of components due chemical reaction in a flow, the intensity of polymer stretching due to the flow, and other transport phenomena problems, cannot be observer-dependent. Hence, if a vortex is non-objective, the logic of cause-effect that could link the flow character with one of these measurers of the intensity that a certain phenomenon is occurring is weaker, when compared to an objective criterion. 

In the present work we employ objective versions of four classic criteria largely used in the literature. The classic criteria and their respective objective versions are analysed and applied for two benchmark cases, the transient {\bf Arnold-Beltrami-Childress (ABC) flow \cite{Arnold-65,Childress-67a,Childress-67b} and the flow through a 4:1 contraction}.

\section{Classic criteria}\label{sec:classiccriteria}
In the following we briefly present four criteria that are currently used in the literature to classify different regions of the flow. These criteria are Eulerian and Galilean-invariant and were recently selected by Pierce \etal \cite{Pierce-13} to evaluate for instance boundary layer flows.

The $\q$-criterion was proposed by Hunt \etal \cite{Hunt-88} in the context of incompressible flows. Besides local pressure minima, they required that, to identify a vortex, the second invariant of the velocity gradient tensor, $\bgradu$ (defined by $\bgradu = (\partial u_j/\partial x_i) \boldsymbol{e}_i\boldsymbol{e}_j$, where $\boldsymbol{u}=u_i\boldsymbol{e}_i$ is the velocity vector field), should be positive. This condition can be expressed for incompressible flows as a function of the Euclidean norms\footnote{The Euclidean norm of a generic second order tensor $\boldsymbol{A}$ is $\Vert \boldsymbol{A} \Vert = \sqrt{\textnormal{tr}(\boldsymbol{A} \cdot \boldsymbol{A}^T)}$, where $\textnormal{tr}(\cdot)$ is the first invariant (trace) of a given second order tensor.} of the symmetric, $\bD = (\bL + \bL^T)/2$, and skew symmetric, $\bW = (\bL - \bL^T)/2$, parts of the velocity gradient (where $\bL = (\partial u_i/\partial x_j) \boldsymbol{e}_i\boldsymbol{e}_j$ is the transpose of $\bgradu$). The condition for the $\q$-criterion can be expressed as follows,

\begin{equation}
Q = \frac{1}{2} \left( \Vert \bW \Vert ^2 - \Vert \bD \Vert ^2 \right) > 0 \; .
\label{eq:q}
\end{equation}

The $\d$-criterion proposed by Chong \etal \cite{Chong-90} is based on the assumption of an equivalence between a vortex and complex eigenvalues of the velocity gradient tensor\footnote{The expression \emph{velocity gradient} is used interchangeably for $\bL$ or $\bgradu$.}. Complex eigenvalues of the velocity gradient is a sign of vorticity dominance with respect to rate-of-strain, since the symmetric rate-of-strain tensor can only have real eigenvalues. Mathematically, the $\d$-criterion can be defined as

\begin{equation}
\Delta = \left( \frac{Q}{3} \right)^3 + \left( \frac{\textnormal{det}(\bD+\bW)}{2} \right)^2 > 0  \; ,
\label{eq:d}
\end{equation}
where $\textnormal{det}(\cdot)$ is the third invariant (determinant) of a given second order tensor.

The $\ltwo$-criterion proposed by Jeong and Hussain \cite{Jeong-95} is based on the idea of joining the local pressure minima condition to a vorticity predominance over the rate-of-strain in the same mathematical condition. By making some assumptions, neglecting some terms, this condition leads to 

\begin{equation}
\ltwo = \ltwo^{\bD^2 + \bW^2} < 0  \; ,
\label{eq:l2}
\end{equation}
where $\ltwo^{\bD^2 + \bW^2}$ is the intermediate eigenvalue of the tensor $\bD^2 + \bW^2$.

The $\lcrlci$-criterion proposed by Chakraborty \etal \cite{Chakraborty-05} is based on the concept that material points that follow orbits which remain compact during the revolutions around each other are in a vortex. $\lambda_{ci}$ and $\lambda_{cr}$ are the imaginary and real values, respectively, of the conjugate complex eigenvalues of the velocity gradient: 

%
\begin{equation}
\frac{\lambda_{cr}}{\lambda_{ci}} = \frac{\lambda_{cr}^{\bD+\bW}}{\lambda_{ci}^{\bD+\bW}} = \frac{\lambda_{cr}^{\bL}}{\lambda_{ci}^{\bL}} \; .
\label{eq:lcrlci}
\end{equation}

The criterion was proposed in the form $\lcrlci<\epsilon$ and, therefore,  a threshold parameter ($\epsilon$) is needed. This condition can only be applied in regions where there are complex eigenvalues of the velocity gradient. This criterion was conceived to be applied together with the $\lambda_{ci}>\delta$ criterion proposed by Zhou \etal \cite{Zhou-99}, where another threshold is needed. Since we will use a normalized dimensionless version of the criterion, these two conditions can be condensed into a single one, where to assume that $\lambda_{ci}$ is significant is equivalent than to assume that $\lcrlci$ is small.

\subsection{Objective redefinition for classic criteria}\label{objredef}

Drouot \cite{Drouot-76a}, Drouot and Lucius \cite{Drouot-76b} have shown that the relative rate-of-rotation tensor, $\bWbar$, defined by

\begin{equation}
\bWbar = \bW - \OmegaD \; ,
\label{eq:wbar}
\end{equation}
is an objective quantity. In Eq.~(\ref{eq:wbar}), $\OmegaD$ accounts for the rate-of-rotation of the eigenvectors of $\bD$, and is defined by

\begin{equation}
\mathbf{\OmegaD} \equiv \hat{e}^{\bD}_i \dot{\hat{e}}^{\bD}_i \; ,
\label{eq:omegad}
\end{equation}
where $\hat{e}^{\bD}_i$ are the normalised eigenvectors of $\bD$, and $\dot{\hat{e}}^{\bD}_i$ are their time derivatives. This quantity was used by Astarita \cite{Astarita-79}, to form an index defined by the ratio of the norm of $\bWbar$ to the norm of $\bD$. This index was used as means for flow classification dividing the domain into extension-like motions and rigid-body-rotation-like ones. Astarita \cite{Astarita-79} was seeking for an objective quantity when he proposed that flow classifier. 

Although a different path was followed by Dresselhaus and Tabor \cite{Dresselhaus-92} and Tabor and Klapper \cite{Tabor-94} investigating the alignment and stretching of material filaments\footnote{\emph{Material filaments} being straight lines of infinitesimal size in the fluid which can rotate, stretch, compress, but not bend.} in an approach of dynamic systems, they have also arrived on the necessity of expressing these quantities with the help of the relative-rate-of-rotation tensor, which was called \emph{effective vorticity}. The physical interpretation of the effective vorticity is discussed next. 

The more common interpretation of the vorticity tensor is that a vorticity component associated to a certain plane is the (arithmetic) mean of the rate-of-rotation of two filaments initially orthogonal to each other in that plane. However, a less adopted interpretation is that the vorticity is the rate-of-rotation of the filaments which are aligned to the eigenvectors of $\bD$. This fact induces the necessity of evaluating the vorticity with respect to the rate-of-rotation of the eigenvectors of $\bD$. When the eigenvectors of $\bD$ are fixed in certain frame of reference, the vorticity tensor is responsible for a deviation between the filaments that are initially aligned with the eigenvectors and their respective directions. However, when the flow is complex and the eigendirections of $\bD$ do change in time, the effective vorticity is the entity responsible for this deviation and, by consequence, for exposing different material filaments to the eigenvectors. This also means that when effective vorticity vanishes, the same material filament is exposed to the eigendirections of the rate-of-strain tensor. In this case, the filament aligned to the eigenvector corresponding to the highest positive eigenvalue is persistently stretched. As shown by Dresselhaus and Tabor \cite{Dresselhaus-92}, effective vorticity plays an important role on the dynamics of material lines and vorticity lines, affecting vortex stretching also.

As suggested by Haller \cite{Haller-05}, an alternative but analogous form of the $\q$-criterion can be built by replacing vorticity by effective vorticity in Eq.~(\ref{eq:q}) leading to

\begin{equation}
\qobj = \frac{1}{2} \left( \Vert \bWbar \Vert ^2 - \Vert \bD \Vert ^2 \right) > 0 \; .
\label{eq:qobj}
\end{equation}
A similar procedure can be adopted for adapting the other criteria into an objective backbone. Therefore we can define the new versions of $\d$-, $\ltwo$-, and $\lcrlci$-criteria, respectively, as
\begin{equation}
\dobj = \left( \frac{\qobj}{3} \right)^3 + \left( \frac{\textnormal{det}(\bD+\bWbar)}{2} \right)^2 > 0 \; ,
\label{eq:dobj}
\end{equation}
\begin{equation}
\ltwoobj = \lambda_2^{\bD^2 + \bWbar^2} < 0 \; ,
\label{eq:l2obj}
\end{equation}
\begin{equation}
\frac{\hat{\lambda}_{cr}}{\hat{\lambda}_{ci}} = \frac{\lambda_{cr}^{\bD+\bWbar}}{\lambda_{ci}^{\bD+\bWbar}} = \frac{\lambda_{cr}^{\bLbar}}{\lambda_{ci}^{\bLbar}} \; .
\label{eq:lcrlciobj}
\end{equation}

We highlight that all objective entities hereafter will be displayed with a hat.

\section{Results}\label{sec:resu}
This section presents the results obtained by applying both objective and non-objective criteria to two cases: (1) the three-dimensional analytical flow field known as the ABC flow in its unsteady version, and (2) an abrupt 4:1 planar contraction.

It is worth noting that all criteria have been normalized in order to produce values between 0 and 1. Moreover, normalized values greater than or equal to 0 and less than 0.5 represent swirling-like or elliptical regions, whereas those greater than 0.5 and less than or equal to 1 represent non-swirling-like or hyperbolic regions. The value of 0.5 represents then a transition (parabolic) region where the magnitude of rotation rate and deformation rate are alike. The only exception applies to the $\lcrlci$- and $\lcrlciobj$-criteria. Since these criteria apply for elliptical regions only, we decided to normalize them from 0 to 0.5, where 0.5 is the boundary of elliptical region. 

The difference from the non-objective and objective versions lies on {\bf the} reference frame from which the rate-of-rotation is computed. While the non-objective quantities use the original fixed frame, the objective quantities have their reference on the local frame attached to the eigenvectors of $\bD$.  


Normalized criteria are identified by a superscript star and are given by the following equations \cite{Thompson-09}

\begin{equation}
 \qn = \dfrac{1}{\pi} \cos^{-1} \left({\dfrac{\Vert{\bW\Vert}^2 - \Vert{\bD}\Vert^2}{\Vert{\bW}\Vert^2 + \Vert{\bD}\Vert^2}  } \right) \; ,
\label{eq:Qnorm}
\end{equation}

\begin{equation}
 \dn = \dfrac{1}{\pi} \cos^{-1} \left[{\dfrac{\left(\dfrac{{\Vert{\bW}\Vert^2 - \Vert{\bD}\Vert^2}}{6} \right)^3 + \left(\dfrac{\textnormal{det}(\bD+\bW)}{2} \right)^2}{\left(\dfrac{{\Vert{\bW}\Vert^2 + \Vert{\bD}\Vert^2}}{6} \right)^3 + \left(\dfrac{\textnormal{det}(\bD+\bW)}{2} \right)^2}}\right] \; , 
\label{eq:dnorm}
\end{equation}
\begin{equation}
 \ltwon = 1 - \dfrac{1}{\pi} \cos^{-1} \left[{\dfrac{{\lambda_2}^{\bD^2 + \bW^2}}{tr(\bD^2) - {\lambda_1}^{\bD^2 + \bW^2} - {\lambda_3}^{\bD^2 + \bW^2}}}\right] \; , 
\label{eq:l2norm}
\end{equation}
\begin{equation}
 \frac{\lambda_{cr}^*}{\lambda_{ci}^*} = \dfrac{2}{\pi^2} \left[{\tan^{-1}\left({\dfrac{{\lambda_{cr}}^{\mathbf{\bD+\bW}}}{{\lambda_{ci}}^{\mathbf{\bD+\bW}}}}\right)}\right]^2 \; ,
\label{eq:lcrcinorm}
\end{equation}
\begin{equation}
 \qobjn = \dfrac{1}{\pi} \cos^{-1} \left({\dfrac{\Vert{\bWbar\Vert}^2 - \Vert{\bD}\Vert^2}{\Vert{\bWbar}\Vert^2 + \Vert{\bD}\Vert^2}  } \right) \; ,
\label{eq:Q-objnorm}
\end{equation}
\begin{equation}
 \dobjn = \dfrac{1}{\pi} \cos^{-1} \left[{\dfrac{\left(\dfrac{{\Vert{\bWbar}\Vert^2 - \Vert{\bD}\Vert^2}}{6} \right)^3 + \left(\dfrac{\textnormal{det}(\bD+\bWbar)}{2} \right)^2}{\left(\dfrac{{\Vert{\bWbar}\Vert^2 + \Vert{\bD}\Vert^2}}{6} \right)^3 + \left(\dfrac{\textnormal{det}(\bD+\bWbar)}{2} \right)^2}}\right] \; , 
\label{eq:d-objnorm}
\end{equation}
\begin{equation}
 \ltwoobjn = 1 - \dfrac{1}{\pi} \cos^{-1} \left[{\dfrac{{\lambda_2}^{\bD^2 + \bWbar^2}}{tr(\bD^2) - {\lambda_1}^{\bD^2 + \bWbar^2} - {\lambda_3}^{\bD^2 + \bWbar^2}}}\right] \; , 
\label{eq:l2-objnorm}
\end{equation}
\begin{equation}
 \frac{\hat{\lambda}_{cr}^*}{\hat{\lambda}_{ci}^*} = \dfrac{2}{\pi^2} \left[{\tan^{-1}\left({\dfrac{{\lambda_{cr}}^{\mathbf{\bD+\bWbar}}}{{\lambda_{ci}}^{\mathbf{\bD+\bWbar}}}}\right)}\right]^2 \; .
\label{eq:lcrci-objnorm}
\end{equation}

\subsection{Transient ABC flow}\label{subsec:abc}
The classic ABC flow \cite{Arnold-65,Childress-67a,Childress-67b} is largely used in the study of chaotic trajectories. Aiming to investigate high-frequency instabilities \cite{Friedlander-91,Lifschitz-91}, we used a transient version of the ABC flow, also considered by Haller \cite{Haller-05}. The flow field is given by the following set of equations:

\begin{align}
 u(y,z,t) &= A(t)\sin(z) + B\cos(y) \, , \nonumber \\
 \label{eq:abc}
 v(x,z,t) &= B\sin(x) + A(t)\cos(z) \, , \\  
 w(x,y,t) &= C\sin(y) + B\cos(x)    \, . \nonumber
\end{align}

In the set of equations (\ref{eq:abc}), $A$ is time-dependent and is defined as $A(t) = A_0 + \parens{1+e^{-qt}}\sin(\omega t)$ with $A_0=\sqrt{3}$, $q=0.1$ and $\omega=2\pi$, while $B = \sqrt{2}$ and $C = 1$. 

\begin{figure}[ht]
 \centering
 \includegraphics[width=1.0\textwidth]{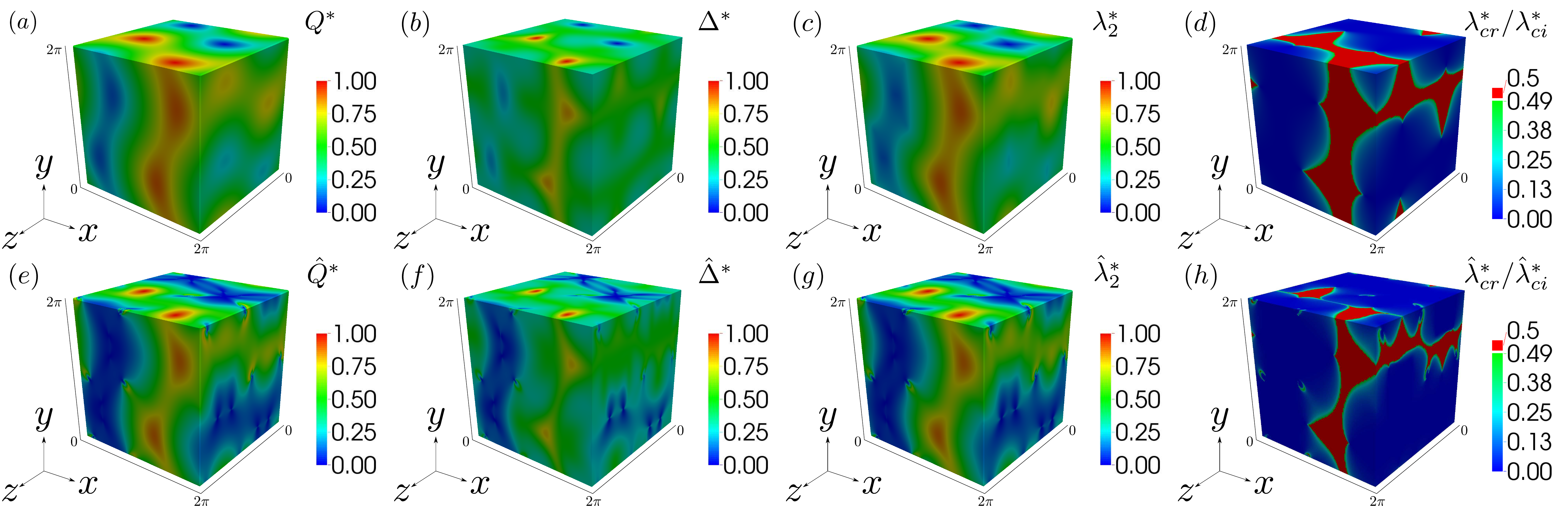}
 \caption{Iso-contours of the normalized flow classification criteria applied to the ABC flow field: non-objective (first line) and objective (second line) versions of $\qn$-criterion, $\dn$-criterion, $\ltwon$-criterion and $\lcrlcin$-criterion.}
 \label{fig:abc}       
\end{figure}

Figure~\ref{fig:abc} contains the normalized classic flow classification criteria ($\qn$, $\dn$, $\ltwon$ and $\lcrlcin$) on the first line and their respective normalized objective version ($\qobjn$, $\dobjn$, $\ltwoobjn$ and $\lcrlciobjn$) on the second line. {\bf The non-objective criteria were obtained using the instantaneous velocity flow field described by the set of equations~(\ref{eq:abc}), whereas the objective criteria need two consecutive velocity fields to compute the rate-of-rotation of the eigenvectors of $\bD$, $\OmegaD$ (see Eq.~(\ref{eq:omegad}) above).} 

The figure shows three surfaces of the cube whose dimensions are limited to the interval $[0,2\pi]$. Generally, the normalized $\q$- and $\ltwo$-criteria (first and third columns in Fig.~\ref{fig:abc}, respectively) present very similar results, which may be explained by the strong relation between these two criteria (see \cite{Jeong-95}). The classifications of the normalized $\d$-criteria (second column in Fig.~\ref{fig:abc}) are also qualitatively similar when compared to those obtained using the normalized $\q$-criteria, although slightly different quantitatively, which, once again, may be justified by the relation between the $\d$-criterion and the $\q$-criterion (\cite{Jeong-95}). Both $\lcrlcin$- and $\lcrlciobjn$-criteria (last column in Fig.~\ref{fig:abc}) have a different range due to less detailed information on non-swirling-like regions. Nevertheless, they maintain a similar qualitative behaviour in terms of the location of vortex cores.

Analysing the non-objective criteria (first line in Fig.~\ref{fig:abc}) at the plane $z=2\pi$, the $\qn$-, $\dn$- and $\ltwon$-criteria (Figs.~\ref{fig:abc}a-c) identify two vortex cores at $(x \approx \pi/2,y \approx \pi/2)$ and $(x \approx \pi/2,y \approx 3\pi/2)$, which are characterized by the colour blue. Once again, due to its different range, almost all the region between $0 < x < \pi$ is blue according to the $\lcrlcin$-criterion (Fig.~\ref{fig:abc}d), characterizing a swirling-like region. Still at this plane, all classic criteria seem to identify extensional (red) regions whose cores are located at approximately $(x = 3\pi/2,y = \pi/2)$ and $(x = 3\pi/2,y = 3\pi/2)$. The regions between elliptical (blue) and hyperbolic (red) regions are transition regions (green) generally referred to as \emph{parabolic}, where the role played by rotation and extension is equivalent. Qualitatively, the results at the plane $y=2\pi$ are very similar to those of the plane $z=2\pi$. Two vortex (blue) cores are now identified at $(x \approx \pi/2,z \approx \pi/2)$ and $(x \approx 3\pi/2,z \approx \pi/2)$, and two extensional (red) regions are located at approximately $(x = \pi/2,z = 3\pi/2)$ and $(x = 3\pi/2,z = 3\pi/2)$. It is worth noting that the magnitude of both swirling- and non-swirling-like regions are slightly greater than those identified at the plane $z=2\pi$. Finally, for the plane $x=2\pi$, the behaviour is yet similar, with vortex cores now identified at $(y \approx \pi/2,z \approx \pi/2)$ and $(y \approx \pi/2,z \approx 3\pi/2)$, and extensional regions centred around $(y \approx 3\pi/2,z \approx \pi/2)$ and $(y \approx 3\pi/2,z \approx 3\pi/2)$. The main difference at this plane is the magnitude of the identified motions, which are reasonably weaker, indicating that the intensities of rotational and extensional motions are close to each other.            

\begin{figure}[ht]
 \centering
 \includegraphics[width=0.8\textwidth]{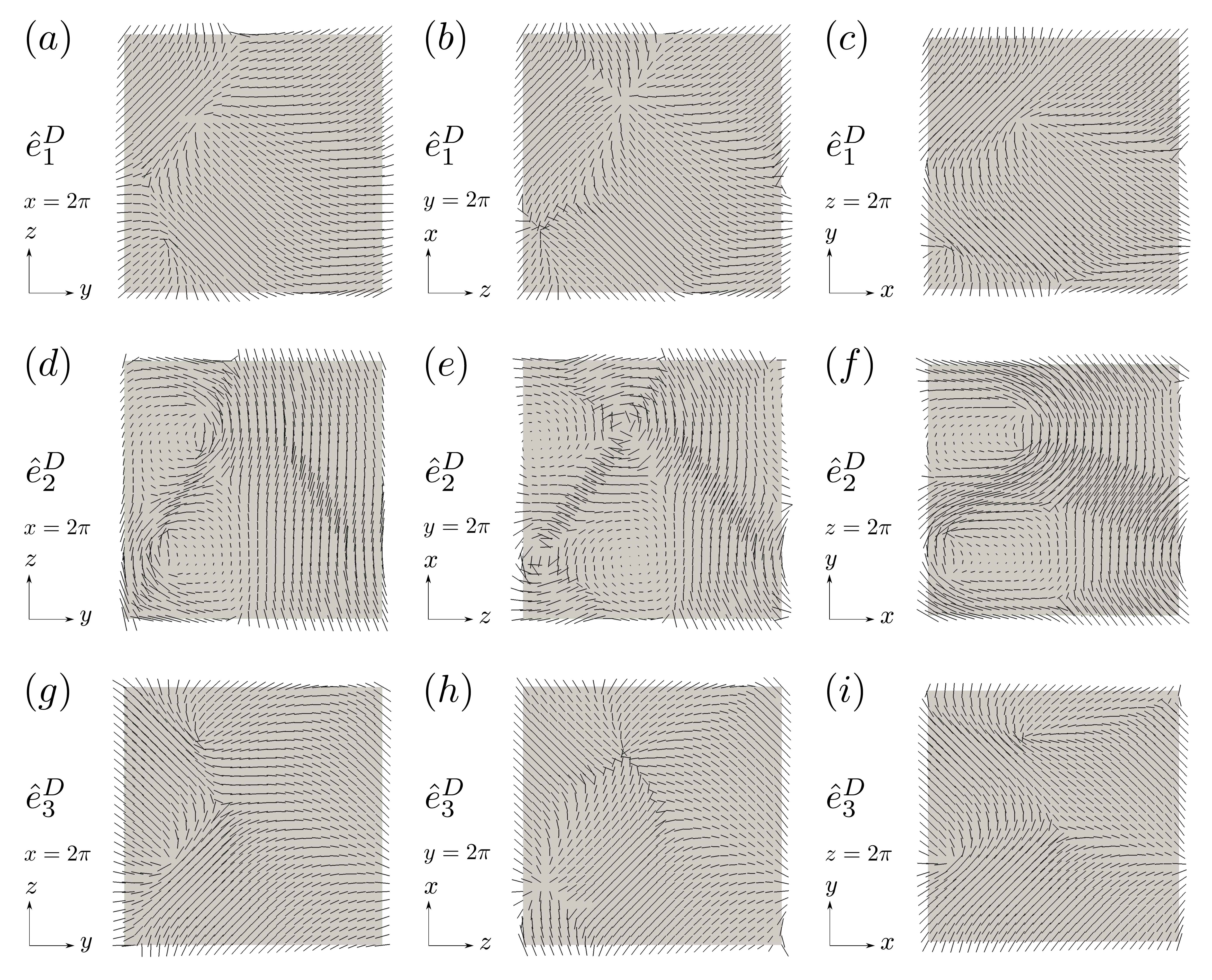}
 \caption{Eigendirections of tensor $\bD$ in the three planes for the ABC flow considered. The ordering corresponds to the eigenvalues $\lambda^{\bD}_1\geq\lambda^{\bD}_2\geq\lambda^{\bD}_3$.}
 \label{fig:abc_eigvec_all}       
\end{figure}

From the perspective of the objective criteria (second line in Fig.~\ref{fig:abc}), the flow at the three planes analysed above presents quite similar characteristics. The main difference is the remarkable increase in swirling-like (blue) regions. This fact is related to regions where the rate of rotation of the eigenvectors of $\bD$ are more pronounced, see Fig.~\ref{fig:abc_eigvec_all}. This means that there are regions where the filaments aligned with the eigenvectors of $\bD$ do not rotate intensively with respect to the reference frame where the problem is being described, but rotate significantly with respect to the frame attached to the eigendirections of $\bD$. This fact will be more explored in the following section for the 4:1 abrupt contraction below.

\subsection{Abrupt 4:1 contraction}\label{subsec:contra}
The flow generated by an abrupt 4:1 contraction presents different regions in which the fluid is submitted to shear, extension or rigid-body motion. Because this feature can be considerably useful for a rich discussion about flow classification, we considered a laminar {\bf steady state} flow of a Newtonian fluid (Reynolds number based on the outlet velocity and height equal to $0.043$). The numerical approach used is the same described by Mompean \etal \cite{Mompean-03} and the mesh consists of 150$\times$80 grid points respectively in the streamwise ($x$) and wall-normal ($y$) directions. 

\begin{figure}[ht]
 \centering
 \includegraphics[width=1.0\textwidth]{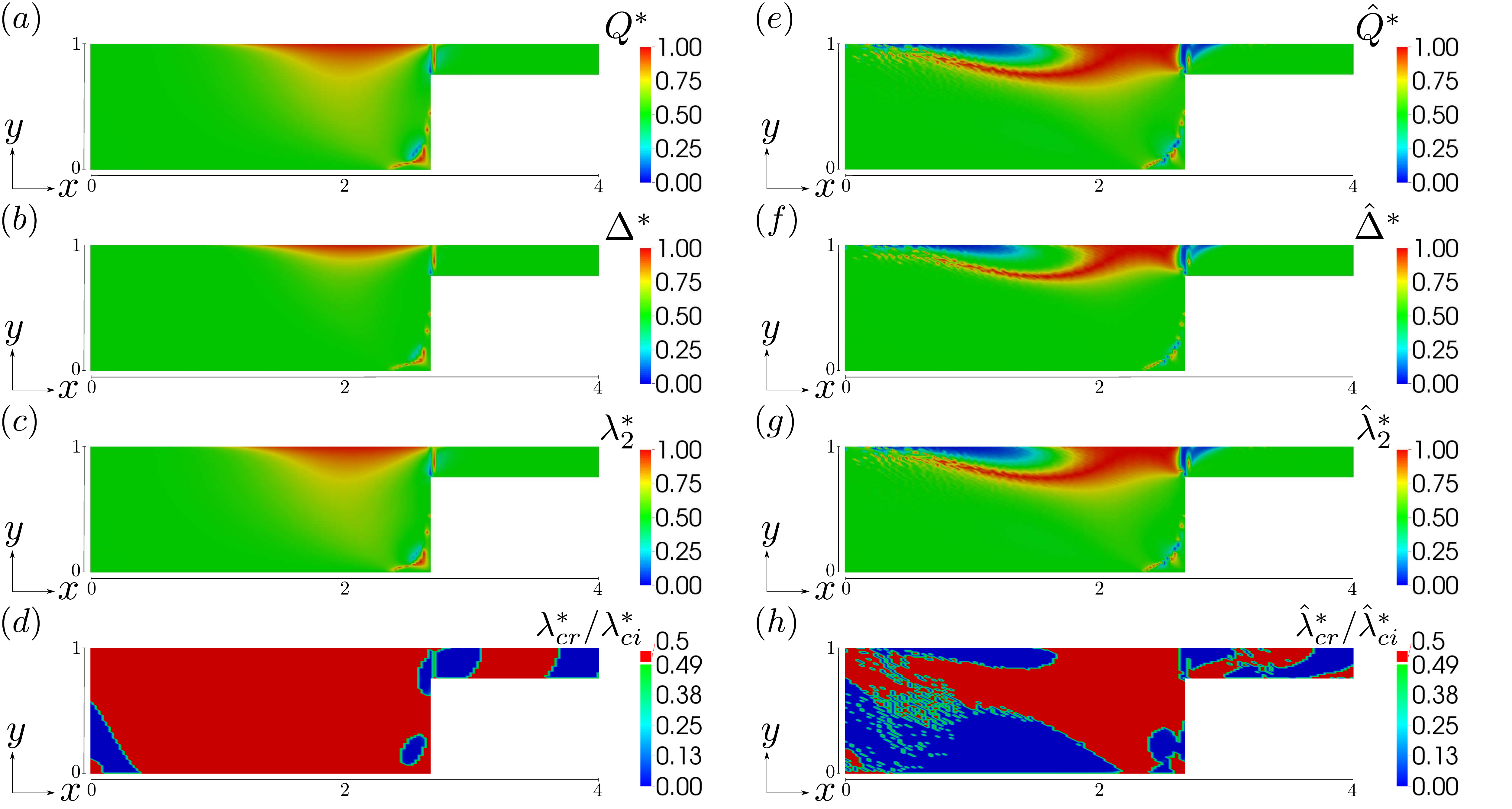}
 \caption{Iso-contours of flow classifiers for the flow trough a 4:1 contraction: non-objective (left-hand side) and objective (right-hand side) versions of $\qn$-criterion, $\dn$-criterion, $\ltwon$-criterion and $\lcrlcin$-criterion.}
 \label{fig:contra}       
\end{figure}

The iso-contours of non-objective (left column) and objective (right column) flow classification criteria are shown in Fig.~\ref{fig:contra}. Likewise the ABC flow, the non-objective criteria employed here provide similar identifications and classifications except for the $\lcrlcin$-criterion, since it has a different range associated to elliptical and parabolic domains only. The same rationale applies to the objective versions.

Again, the green color represents regions which are not elliptical nor hyperbolic. In the present case they are related to the shear motion typical of fully-developed flow in a geometry of constant cross sectional area. 

The non-objective criteria $\qn$, $\dn$ and $\ltwon$ (Figs.~\ref{fig:contra}a-c) identify a shear region away from the contraction, where the flow is fully developed. Close to the contraction and around the symmetry plane, an extensional region can be seen due the acceleration of the fluid passing through the contraction. Both near the sharp corner ($x \approx 2.75$,$ 0.75 \lessapprox y \lessapprox 1$) and the corner vortex ($x \approx 2.5$,$ 0.1 \lessapprox y \lessapprox 0.25$) a mix of extensional and rotational motions is observed. In fact, in these regions, the fluid is submitted to both extensions and rotations caused by the singularity of the geometry nearby. It is worth remembering that the $\lcrlcin$-criterion does not provide any distinction in terms of non-swirling-like motions, i.e. swirling-like motions have a degree of intensity (greater than or equal to 0 and less than 0.5) whereas non-swirling-like motions possess only the classification (equal to 0.5, without a gradation). Therefore, in Fig.~\ref{fig:contra}d, every region in red may be interpreted simply as a non-swirling-like region (possibly being either shear or extensional). The blue regions, on the other hand, can be considered as elliptical regions. Thus, except for the size and intensity, the same two regions with rotational motions pointed by the other non-objective criteria are identified by the $\lcrlcin$-criterion. However, it also identifies two extra swirling regions: (a) the one close to the wall in the inner region, and (b) the one just by the outlet region.  

\begin{figure}[ht]
 \centering
 \includegraphics[width=1.0\textwidth]{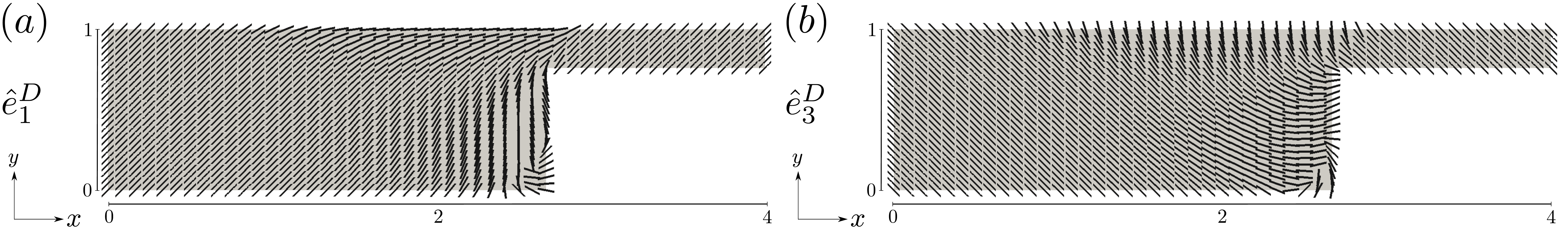}
 \caption{Eigendirections of tensor $\bD$ in the $x-y$ plane for the 4:1 contraction flow. Again, the ordering corresponds to the eigenvalues $\lambda^{\bD}_1\geq\lambda^{\bD}_2\geq\lambda^{\bD}_3$.}
 \label{fig:contraction_eigvec_all}       
\end{figure}

Concerning the objective criteria (Figs.~\ref{fig:contra}e-h), the same three elliptical (swirling-like) regions identified by Mompean \etal \cite{Mompean-03} are observed here. As commented by the authors, the regions (i) just after the sharp corner ($x \approx 2.75$,$ 0.75 \lessapprox y \lessapprox 1$), and (ii) near the corner vortex ($x \approx 2.5$,$ 0.1 \lessapprox y \lessapprox 0.25$) are intuitively understandable, since the flow is submitted to rotate at rates which are larger than the local rate-of-deformation. On the other hand, the third region (iii) just before the extensional region may be counter-intuitive at first, since it appears between regions of expected shear and extension. Again, the authors explain that this region should actually be considered as a plug flow (close to rigid-body motion) where the rates-of-rotation of the eigenvectors of $\bD$ are larger than the local rates-of-deformation. One important result shown by the objective quantities is the information with a broad-spectrum
of values for all the four criteria when compared with the non-objective quantities. This information is frame-invariant and related to the principal directions of the rate-of-deformation tensor, and is physically expected. Before entering a region where extension is dominant, the region just before the contraction plane, the eigenvectors of $\bD$ rotate from the $\pi/4-3\pi/4$ directions of the shear flow in the fully-developed region to the $0-\pi/2$ directions of the extensional flow. This rotation is overlooked by the non-objective criteria, but are captured by the objective ones. Although the filaments do not rotate predominantly with respect to the reference frame $(x,y)$, they do rotate with respect to the eigenvectors of $\bD$, as clearly shown in Fig.~\ref{fig:contraction_eigvec_all}. Because of this relative rate-of-rotation, new filaments are exposed to the corresponding eigenvalues. A similar rationale explains the appearance of the elliptical region just after the contraction, not detected by the non-objective quantities. The $\bD$-eigenvectors need to rearrange in order to return to a fully-developed condition.

It seems that the $\lcrlci$-criterion has some problems on the so-called \emph{parabolic regions}, i.e. the regions of transition between hyperbolic and elliptical regions. It is difficult to delineate this intermediate region by this criterion in both, objective and non-objective versions. We can notice an almost binary color result. What we can see is that parts of the domain with no apparent physical difference that are classified as hyperbolic or elliptical. Most of these parts are in the parabolic region accordingly to other criteria.

\section{Final remarks}\label{sec:final}
In the present work we analysed the performance of normalized objective versions of classic flow classification criteria. The classic criteria are the $\q$-criterion proposed by Hunt \etal \cite{Hunt-88}, the $\d$-criterion proposed by Chong \etal \cite{Chong-90}, the $\ltwo$-criterion proposed by Jeong and Hussain \cite{Jeong-95}, and the $\lcrlci$-criterion proposed by Chakraborty \etal \cite{Chakraborty-05}. The two flows considered were the transient ABC flow with $A(t) = A_0 + \parens{1+e^{-qt}}\sin(\omega t)$ (with $A_0=\sqrt{3}$, $q=0.1$ and $\omega=2\pi$), $B = \sqrt{2}$ and $C = 1$, and a 4:1 abrupt contraction. The ABC flow is known for its chaotic character for the values employed. The abrupt contraction is a complex flow that exhibits different types of motion distributed in the domain: extension, pure shear, and rigid-body motion. 

The objective quantities in the present work use the effective vorticity as the entity associated to the elliptical character of the flow, in the place of the vorticity. The main difference between objective and non-objective quantities was the presence of elliptical regions in the objective versions which were not present in its non-objective counterparts.  This happened because in these regions, the rate-of-rotation of the eigenvectors of the rate-of-strain tensor, with respect to the original frame, was significant and the effective vorticity uses this rate-of-rotation as reference for computing the rate-of-rotation of the filaments.


Finally, it was shown that the objective criteria provides more information about the kinematics of the flow. 
Such new features of the flow may be useful in the investigation of complex flows and phenomena, as, for instance, drag-reducing flows, convection-driven problems or the mixture between two or more fluids. {\bf We emphasize that our goal is not to appoint to a preferable particular flow classification criterion, but to stand up for the advantages of applying criteria which enjoy objectivity}.



\section*{Acknowledgements}
The authors would like to express their acknowledgement and gratitude to the Brazilian Scholarship Program \emph{Science Without Borders}, managed by CNPq (National Council for Scientific and Technological Development), for the partial financial support for this research.

\end{document}